\documentstyle[manuscript,aps,prc,tighten,amssymb,graphicx]{revtex}
\def\be{\begin{equation}}
\def\ee{\end{equation}}
\def\bea{\begin{eqnarray}}
\def\eea{\end{eqnarray}}
\def\ra{\rightarrow}
\def\cd{\cdot}
\def\thetaf{\mathop{\theta}\nolimits}

\def\fpi{\left|F_\pi(q^2)\right|^2}
\def\om{\omega}
\def\qv{{\bbox{q}}}
\def\zv{{\bbox{0}}}

\begin{document}
\draft

\title{Pion broadening and low-mass dilepton production}

\author{H.-J. Schulze}
\address{Sezione INFN, Dipartimento di Fisica, Universita di Catania,\\
         Via Santa Sofia 64, I-95123 Catania, Italy}

\author{D. Blaschke}
\address{Fachbereich Physik, Universit\"at Rostock,
         Universit\"atsplatz 1, D-18051 Rostock, Germany\\
         \centerline{and}
         Bogoliubov Laboratory for Theoretical Physics,
         JINR Dubna, 141980 Dubna, Russia}

\maketitle
\date{}
\begin{abstract}
 We determine mass and transverse momentum spectra of dileptons produced
 in Pb+Au (158 GeV/u) collisions within a pion annihilation model.
 A fit to the data requires simultaneous mass reduction and broadening
 of the in-medium rho propagator.
 The introduction of a finite pion width, as required within self-consistent
 approaches to the interacting pion gas, further improves the agreement with
 the data.
\end{abstract}
\vskip1.5cm


\section{Introduction}

Since a couple of years the experimental study of dilepton production in
relativistic heavy ion collisions is becoming more and more detailed.
Besides the usual mass spectrum, the latest results of the CERES collaboration
comprise also information on the transverse momenta of the produced dileptons
in Pb+Au (158 GeV/u) collisions \cite{ceres95,ceres96}.

Since several hundreds of pions are produced in such a collision, it is
clear that pion annihilation during the hot and dense initial phase
plays a dominant role for low-mass dilepton production,
apart from the known contributions due to single meson Dalitz decays
after freeze-out.
From the previous analyses it is indeed well known that a major part
of the observed dilepton ``excess'' can be attributed to pion-pion annihilation
in the dense medium that is created during the collision.
However, the experimental electron pair mass spectrum can only be explained by
assuming a modification of the rho meson mass and/or width in the
medium \cite{li,soll,cass,rapp,weise,sriva,old}.

Apart from the knowledge of in-medium meson properties, the theoretical models
require assumptions about the space-time development of the collision.
Usually rather elaborate transport simulations or hydrodynamical models are
used \cite{li,soll,cass}.
A much cruder, but very transparent way is to employ an idealized standard
scenario,
allowing to test easily some qualitative assumptions about the dilepton
production mechanism.
In this article we confront the latest experimental results to theoretical
predictions of such a ``standard model'' of the space-time evolution of an
ultrarelativistic heavy ion collision, namely the Bjorken picture of a
purely longitudinal expansion \cite{sriva,old,bjorken},
involving an initial quark-gluon plasma (QGP) phase,
an intermediate mixed phase,
and a final hadronic (pion gas) phase.

This model involves essentially four parameters, namely the temperatures
$T_i$, $T_c$, and $T_f$, specifying the initial, critical, and freeze-out
temperatures;
as well as the ratio $r$ of degrees of freedom in the quark and
hadronic phase, respectively.
As in Ref.~\cite{old}, we use $r=37/3$
(two flavor quark-gluon plasma versus pion gas)
and the temperatures $T_i=250\;\rm MeV$, $T_c=160\;\rm MeV$,
whereas we fix $T_f=140 \;\rm MeV$,
between the extreme cases of 120 MeV and 150 MeV that were
discussed in Ref.~\cite{old}.
The precise values of these parameters are perhaps not so important as other
fundamental assumptions of the model, such as purely longitudinal expansion
and the idealization of pure quark and pion phases.
We feel, however, that all these factors mainly influence the overall
normalization of the resulting dilepton spectrum and not so much its spectral
shape, which is of major interest in the following theoretical analysis.

\section{Formalism}

We begin with a short review of the relevant expressions.
The total dilepton production rate due to pion annihilation
in lowest order of the electromagnetic interaction can be written as
\bea
 {dN_{l^+l^-}\over d^4x}(x) &=&
 \int_1\int_2\int_3\int_4 f^+(p_1,x)f^-(p_2,x) \;
 (2\pi)^4 \delta^4(p_1+p_2-p_3-p_4) \; |M_{\rm fi}|^2
\eea
with
\be
  \int_i \equiv \int {d^3 p_i \over (2\pi)^3 2E_i}
          = {1\over (2\pi)^3} \int d^4 p_i \, \thetaf(p_i^0)\delta(p_i^2-m_i^2)
\ee
and the thermal pion distributions
\be
 f^\pm (p,x) = {1\over \exp{[p\cd u(x)/T(x)]} - 1} \:,
\ee
where $T(x)$ is the local temperature in the fluid element moving with
four velocity $u(x)$.
The matrix element
\be
 |M_{\rm fi}|^2 = l_{\mu\nu}(p_3,p_4) \,{e^4\over q^4}\, h^{\mu\nu}(p_1,p_2)
 \ , \quad
 q = p_1+p_2 = p_3+p_4
\ee
can be specified in terms of
the hadronic (initial state) and leptonic (final state) tensors,
\bea
 h^{\mu\nu}(p_1,p_2)
 &=& -\fpi (p_1-p_2)^\mu (p_1-p_2)^\nu \:,
\\
 l^{\mu\nu}(p_3,p_4)
 &=& 4 \left[ p_3^\mu p_4^\nu + p_4^\mu p_3^\nu -
 (p_3\cd p_4 + m_l^2)g^{\mu\nu} \right] \:,
\eea
where $m_l$ is the lepton mass and
$F_\pi$ is the experimentally known pion form factor.
At this stage the vacuum matrix element is used, without any modification
due to medium effects.
Defining
\bea
 H^{\mu\nu}(q,x) &\equiv& \int_1\int_2 (2\pi)^4 \delta^4(p_1+p_2 - q)
 \, h^{\mu\nu}(p_1,p_2) \, f^+(p_1,x)f^-(p_2,x) \:,
\\
 L_{\mu\nu}(q) &\equiv& \int_3\int_4 (2\pi)^4 \delta^4(p_3+p_4 - q) \;
 l_{\mu\nu}(p_3,p_4)
\\
 &=& \left( q^2g_{\mu\nu} - q_\mu q_\nu \right) {L(q^2) \over 6\pi}
 \ ,\
 L(q^2) \equiv
 \thetaf(q^2-4m_l^2)
 \left( 1+{2m_l^2\over q^2} \right)
 \sqrt{1-{4m_l^2\over q^2}}
 \:,
\eea
the differential rate for the production of dileptons with four-momentum $q$
at the space-time point $x$ can be written as
\bea
 {dN_{l^+l^-}\over d^4x d^4q}(q,x) &=&
 {1\over (2\pi)^4} L_{\mu\nu}(q)\, {e^4\over q^4}\, H^{\mu\nu}(q,x)
\\
 &=& {e^4\over (2\pi)^4} {L(q^2) \over 6\pi } \fpi
 {4k^2\over q^2}
 \int_1\int_2 (2\pi)^4 \delta^4(p_1+p_2 - q) f^+(p_1,x)f^-(p_2,x)
\label{e:dndxdq}
\eea
with
\bea
 {4k^2\over q^2} &=&
 \thetaf[M^2-(m_1+m_2)^2]
 \left( 1-{(m_1+m_2)^2\over M^2} \right)
 \left( 1-{(m_1-m_2)^2 \over M^2} \right) \:,
\eea
where $M^2=q^2$ and
$k$ is the center-of-mass ($\qv=\zv$) momentum
of the two colliding pions with mass-squared $m_i^2 = p_i^2$.

The expected number of lepton pairs per collision is then:
\bea
 {dN_{l^+l^-}\over d^4q}(q)
 &=& {e^4\over (2\pi)^4} {L(q^2) \over 6\pi} \fpi
 {4k^2\over q^2}
\nonumber\\ && \times
 \int_1\int_2 (2\pi)^4 \delta^4(p_1+p_2 - q)
 \int d^4x \, f^+(p_1,x)f^-(p_2,x) \:.
\label{e:dndq}
\eea
Therefore the problem reduces to the evaluation of the integral of the
distribution functions over space time.
Within the Boltzmann approximation
($f^\pm (p,x) \approx \exp{[-p\cd u(x)/T(x)]}$, valid for $M\gg T$)
these integrations can even be performed analytically \cite{sriva,thmod},
leading to
\bea
 {dN_{l^+l^-}\over d^4q}(q)
 &=& {e^4\over (2\pi)^4} {L(q^2)\over 6\pi} \fpi {H(q^2)\over 8\pi}
 \!\intop\limits_{{\rm pion \atop phase}}\!\!\! d^4x\, e^{-q\cd u(x)/T(x)}
\label{e:dndqh}
\eea
with
\bea
 H(q^2) &=&
 \left( {4k^2\over q^2} \right)^{3/2}
\\
 &=& \thetaf(q^2-4m_\pi^2) \left( 1 - {4m_\pi^2\over q^2} \right)^{3/2}
 \ , \quad (m_1^2=m_2^2=m_\pi^2)
\label{e:h}
\eea
and
\bea
 \intop\limits_{{\rm pion \atop phase}}\!\!\! d^4x\, e^{-q\cd u(x)/T(x)}
 &=&
 6\pi R^2 \tau_c^2
 \left( {r(r-1)\over6} K_0(z_c) +
 {r^2 \over z_c^{6}}\left[ K(z_c)-K(z_f) \right] \right)
 \:,
\eea
where
$R$ is the effective radius of the reaction zone,
$\tau_c$ is the proper time corresponding to the beginning of the
phase transition,
$M_T=\sqrt{M^2+q_T^2}$ is the transverse mass of the dilepton,
$z=M_T/T$,
$K(z) = (z^2+8)z^3 K_3(z)$, and $K_0$ and $K_3$ are Bessel functions.
The overall normalization is fixed by the condition for an
isentropic expansion,
$\pi R^2 \tau T^3 = \kappa\, dN_\pi/dy$,
with $\kappa\approx 0.22$ for an $SU_F$(2) plasma.
One arrives finally at the dilepton spectrum due to pion annihilation:
\bea
 {dN_{l^+l^-}\over dM dq_T dy}(M,q_T) &=&
 {\alpha^2\over 8\pi^4} \left({\kappa\, dN_\pi/dy \over R}\right)^2
 { M q_T \over M_T^6 } L(q^2) H(q^2) \fpi
\nonumber\\ && \times
 \left( {r(r-1)\over6}z_c^6 K_0(z_c) + r^2 \left[ K(z_c)-K(z_f) \right] \right)
 \:.
\label{e:dndm}
\eea
Keeping, however, the Bose distributions in Eq.~(\ref{e:dndq}),
or in any case for a temperature dependent form factor,
the space-time integrations have to be carried out numerically,
as described in detail in Ref.~\cite{old}.
In this case
one can make use of the analytical result for
the phase-space integral appearing in Eq.~(\ref{e:dndxdq}):
\bea
 &&\int_1\int_2 (2\pi)^4 \delta^4(p_1+p_2 - q) f^+(p_1,x)f^-(p_2,x) =
\nonumber\\
 && \hskip50mm
 {T\over 8\pi \bar q}{1\over \exp{(E/T)}-1}
 \ln\left[ { \cosh{(\om_+/T)} - \cosh{(\delta/T)} \over
              \cosh{(\om_-/T)} - \cosh{(\delta/T)}  } \right]
\eea
with
\bea
 \om_\pm &=& {E\over2} \pm {2k\over M} {\bar q \over 2} \:,
\\
 \delta &=& {m_1^2-m_2^2\over M^2} {E\over 2} \:,
\eea
where $E=q\cd u(x)$ is the dilepton energy, and
$\bar q = \sqrt{E^2-M^2}$ is the three-momentum
component of $q$ in the production fluid element moving
with four-velocity $u$.

The previous expressions concern dileptons stemming from the pion phase.
One has to add the contribution due to quark-antiquark annihilation
that we have not written down explicitly \cite{sriva,old,thmod},
because it is not concerned by the considerations in this paper.
In order to compare with the experimental mass and transverse momentum spectra,
the sum of the two parts
has then to be multiplied with the relevant experimental
acceptance function $A(M,q_T)$ and folded with the experimental mass resolution,
see Ref.~\cite{old}.
Finally the experimentally determined contribution due to decays of
mesons produced during the primary nucleon-nucleon collisions
(``cocktail'') has to be added.
We discuss now the results obtained after this procedure.

\section{Results}

\begin{figure}
\includegraphics[totalheight=8.8cm,angle=0,bb=-110 140 -110 680]{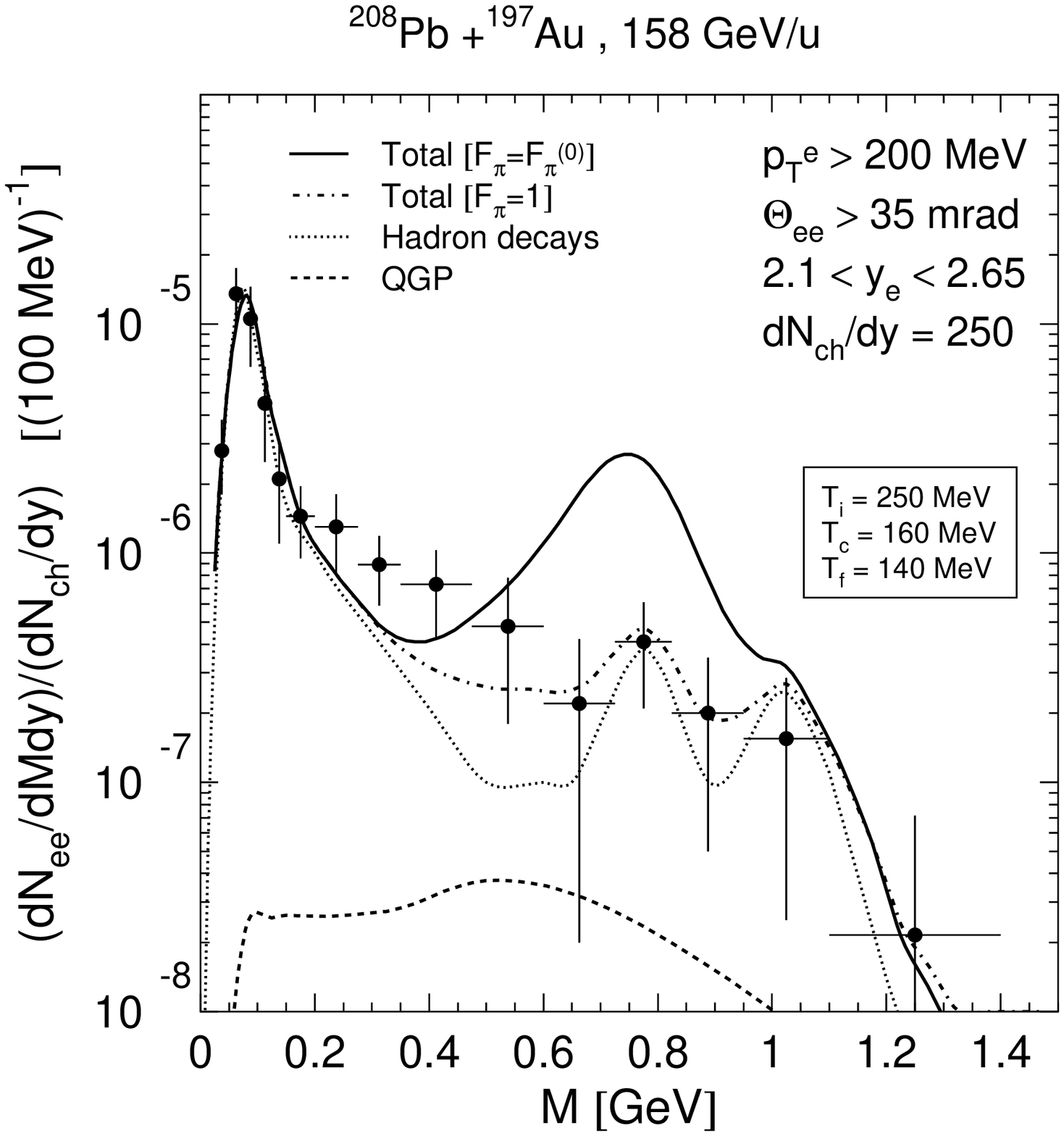}
\caption{
  The dilepton mass spectrum evaluated with a free-space (solid line) or
  a Born (dot-dashed line) pion form factor.
  The theoretical predictions involve acceptance and resolution corrections,
  and the primary meson decay contribution (dotted line) is added.
  The dashed line shows separately the contribution of $q\bar q$ annihilation
  in the quark and mixed phase.
  The data points and the meson decay contribution
  are taken from Ref.~\protect\cite{ceres96}.  }
\label{f:m}
\end{figure}

A first glance at the experimental data
(markers in Fig.~\ref{f:m})
in comparison to the expectation from primary meson decays (dotted line)
shows that
practically all excess dileptons that are produced in the reaction zone
are required to fall within the mass region [200,600] MeV.
In particular close to the bare rho mass ($\approx 800$ MeV)
practically no additional contribution is required.
This fact indicates already a substantial modification of the
in-medium pion form factor,
which is indeed confirmed by our theoretical result using the vacuum pion
form factor (solid curve), which largely overshoots the experimental results
in the rho mass region.
An evaluation using the pure Born cross section, i.e.,
setting tentatively $F_\pi=1$ (dot-dashed curve) yields instead a fair
agreement in the rho mass region with, however, missing strength at lower mass.
It is therefore clear that necessarily
a substantial reduction of the rho meson mass in the medium
is required in order to reproduce the experimental data.

\begin{figure}
\includegraphics[totalheight=8.8cm,angle=0,bb=-110 140 -110 680]{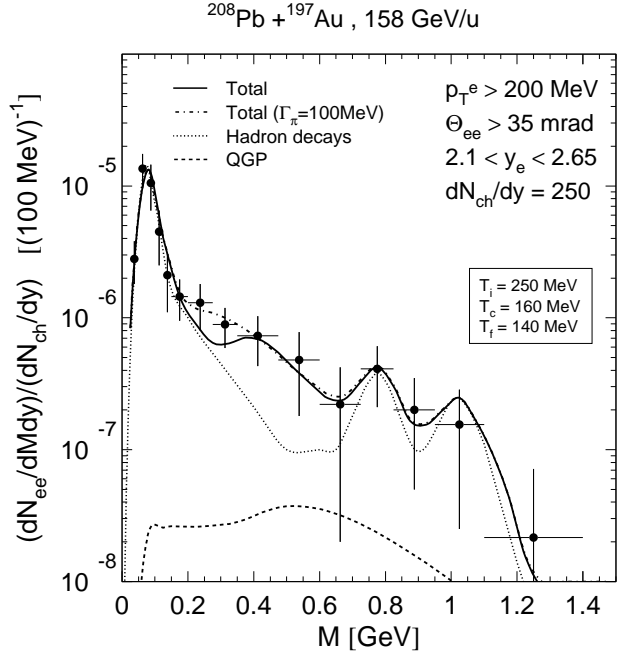}
\caption{
  As Fig.~1, but with the in-medium pion form factor (solid line),
  and additionally assuming a finite pion width (dot-dashed line). }
\label{f:min}
\end{figure}

\begin{figure}
\includegraphics[totalheight=9.1cm,angle=0,bb=-110 140 -110 680]{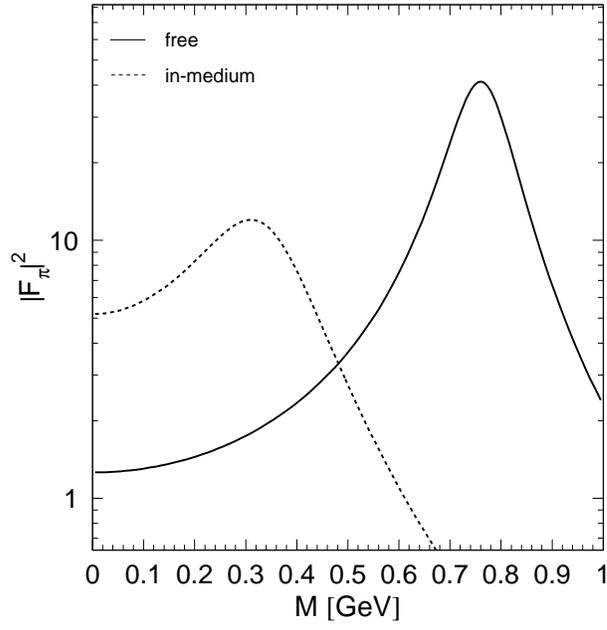}
\caption{
 Free-space and in-medium pion form factor. }
\label{f:fpi}
\end{figure}

Since in our model all excess dileptons stem from a pion gas phase with a small
variation of temperature between $T_c=160$ and $T_f=140$ MeV,
we neglect even the detailed dependence on temperature
and rather attempt to construct an effective in-medium pion form factor
of the standard functional form
\begin{equation}
 \left| F_\pi(M) \right|^2 = { C\, m_\rho^4  \over
 (M^2-m_\rho^2)^2 + m_\rho^2 \Gamma_\rho^2 }
\end{equation}
that allows to reproduce the experimental data.
The constants $m_\rho$, $\Gamma_\rho$, and $C$ are now considered as
free parameters and allowed to deviate from their vacuum values
$m_\rho^0=0.76$ GeV, $\Gamma_\rho^0=0.135$ GeV, and $C^0=1.3$.
Indeed we find that a choice of
$m_\rho = 0.4 m_\rho^0$, $\Gamma_\rho=2\Gamma_\rho^0$, and $C = 7C^0$
allows a very reasonable fit of the data,
shown in Fig.~\ref{f:min} (solid line).
Thus a simultaneous modification of centroid
(reduction by a factor $2\ldots3$)
and width (increase by about a factor 2)
of the pion form factor is required;
we found that changing only one of them does not yield satisfactory results.
The comparison of the resulting effective in-medium and the
vacuum pion form factor is shown in Fig.~\ref{f:fpi}.
We remark at this point that in the present model the contribution of
dileptons from the quark-gluon plasma
(produced during the initial phase and the mixed phase)
is negligible \cite{weise,sriva,huovinen,thoma},
as shown by the dashed line in Figs.~\ref{f:m} and \ref{f:min}.
This is mainly due to the short duration of this phase compared to the
pion phase.

\begin{figure}
\includegraphics[totalheight=9.1cm,angle=0,bb=-110 140 -110 680]{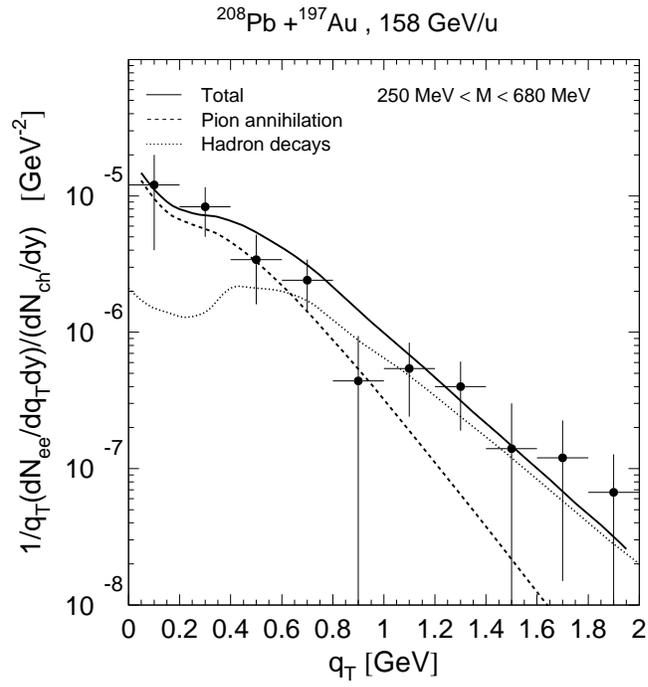}
\caption{
  The dilepton transverse momentum spectrum. }
\label{f:qt}
\end{figure}

After discussing the mass spectrum of electron pairs, we come now to
the distribution of transverse momenta.
Fig.~\ref{f:qt} shows the experimental values
(in the mass interval of interest $250\;{\rm MeV}<M<680\;{\rm MeV}$)
in comparison with our theoretical prediction (solid line).
The good agreement is indeed remarkable, given the fact that in our model
the transverse momentum of the dileptons is entirely due to the thermal
motion of the colliding pions in the longitudinally expanding medium.
This is the reason for the larger slope of the transverse momentum
distribution of dileptons stemming from pion annhihilation (dashed line)
compared to the Dalitz dileptons (dotted line).
The sum of both results in fair agreement with the data.
It seems thus that the transverse momenta of the dileptons reveal no hint
of a transverse expansion of the reaction zone.

\begin{figure}
\includegraphics[totalheight=9.1cm,angle=0,bb=-110 140 -110 680]{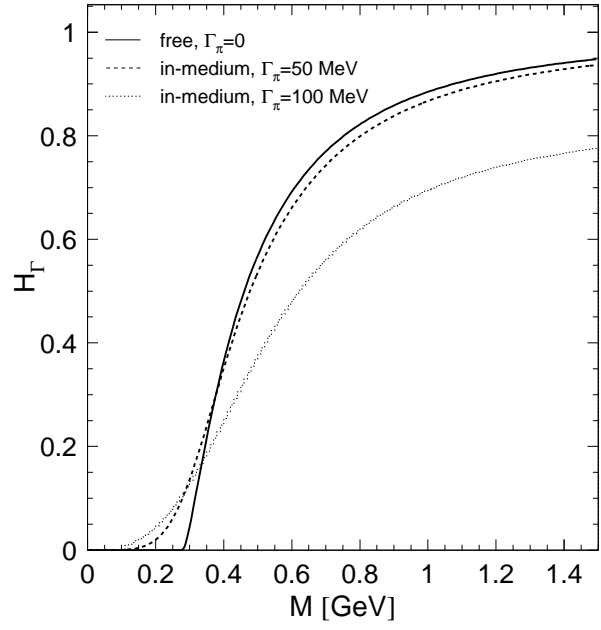}
\caption{
 Kinematic factor, Eq.~(\ref{e:hpi}). }
\label{f:h}
\end{figure}

The in-medium pion annihilation contribution to the dilepton mass spectrum
is naturally cut off at the two-pion threshold, and leaves therefore
a small dip in the theoretical curves around $M\approx 300$ MeV
(solid lines in Figs.~\ref{f:m} and \ref{f:min}).
A natural remedy would be a decrease of the pion mass in the medium, for
which, however, there is so far no solid theoretical foundation.
In this article we rather study the consequences of introducing empirically
a finite damping width of the pion, due to resonant pion rescattering in the
medium.
The occurence of a finite pion damping width  $\Gamma_\pi \approx 50$ MeV
is well known from the study of the $\pi N \Delta$ system \cite{vbrs}
and is also expected to occur in a hot dense pion gas,
due to the strong coupling to the rho and $a_1$ resonances.
Within a simple field theoretical model for this system,
motivated by a quartic self-coupling of the pions, van Hees and Knoll
\cite{hk} have shown how the self-consistent equations of motion can be
derived using Baym's $\Phi$ functional approach \cite{baym}.
As a result of these models, a temperature and momentum dependent pion width
has been obtained \cite{bvy},
which in the relevant domain of $T\approx 150$ MeV
amounts to $\Gamma_\pi \approx 50 \ldots 100$ MeV.
We therefore assume a Gaussian pion mass distribution of width $\Gamma_\pi$,
\be
 g_\pi(m) = \frac{1}{\sqrt{2\pi}\Gamma_\pi}
\exp \left[-\frac{(m-m_\pi)^2}{2\Gamma_\pi^2}\right] \:,
\ee
and naively generalize Eq.~(\ref{e:dndqh})
to the reaction of two pions with different masses
$\pi(m_1) + \pi(m_2) \ra \gamma(q)$.
This is achieved by replacing the function $H(q^2)$, Eq.~(\ref{e:h}), by
\bea
 H_{\Gamma_\pi}(q^2) &=&
 \int_0^\infty\!\! dm_1\, g_\pi(m_1) \int_0^\infty\!\! dm_2\, g_\pi(m_2)
 \thetaf[q^2-(m_1+m_2)^2]
\nonumber\\ && \times
 \left[ \left(1-{(m_1+m_2)^2\over q^2} \right)
 \left( 1-{(m_1-m_2)^2 \over q^2} \right) \right]^{3/2} \:,
\label{e:hpi}
\eea
which is plotted in Fig.~\ref{f:h} for two choices of the pion width
$\Gamma_\pi=50, 100$ MeV.
The resulting dilepton spectrum for a value $\Gamma_\pi=100$ MeV
(and form factor parameters
$m_\rho = m_\rho^0/3$, $\Gamma_\rho = 2\Gamma_\rho^0$, and $C = 40C^0$)
is also shown in Fig.~\ref{f:min} (solid line).
Indeed the introduction of a finite pion width can perfectly describe
the spectrum also in the threshold region $M \approx 2m_\pi$.

Of course, there are other processes that could potentially contribute
dileptons in this mass region around 300 MeV.
In particular, we neglected completely the possible presence of
baryons in the reaction zone and of baryon induced reactions like
$\rho N\ra B$, which might contribute in the low-mass region.
Also possible mesonic rescattering of the rho resonance,
$\rho\pi\rightarrow\pi e_+e_-$,
has been disregarded \cite{baier}.
However, since the mass spectrum below 200 MeV is perfectly reproduced
by $\pi_0$, $\eta$, and $\eta'$ Dalitz decays \cite{eta},
there is not much freedom for such additional ``exotic'' contributions,
since these would dominantly contribute in this low-mass domain,
which is on the contrary not the case for the pion width mechanism
discussed above.
Indeed also theoretical estimates of these processes predict too
small rates \cite{huovinen,pirho}.

\section{Conclusions}

In summary, the latest Pb+Au CERES data hint to a
simultaneous reduction of the rho meson mass and an increase of its width
in the dilepton production medium.
Even if our model (Bjorken scenario)
of the space-time evolution of the collision is rather idealized,
we think that these qualitative conclusions based on the shape of
the dilepton mass distribution can be drawn.
The further ad-hoc introduction of a finite mass distribution of the
interacting pions
served to simulate the effect of a finite pion damping width,
which is required for a consistent treatment of rho mesons as decaying
resonances in an interacting pion gas.
It allowed us to reproduce the experimental results around the two-pion
threshold even better.

Surprisingly, the transverse momenta of the dileptons
were also rather well described by our model, and thus
appear to reveal no hint on a transverse expansion of the reaction zone.
Finally, dilepton production from an eventual QGP phase is negligible due
to the short duration of such a phase,
even if its existence is an assumption of the chosen theoretical
model and influences indirectly, through a consistent time evolution,
also the hadronic rates.

The present investigation can be improved by applying a more realistic
hydrodynamical evolution which goes beyond the simple Bjorken scenario.
The pion broadening should be considered within a definite microscopic model,
so that its temperature and momentum dependence can be included into the
calculation.
Such a model should allow to calculate the pion broadening
consistently together with the rho meson spectral properties.
Since already the simple calculation of the rho meson spectral function
with on-shell pions in a hot pion gas yields a downwards shift of the
rho meson mass by about 30 MeV \cite{barz},
it can be expected that off-shell pions with about $100$ MeV width could
explain an even larger rho meson mass shift, which is inevitable in order to
explain the new low-mass dilepton data of the CERES collaboration.
The objective of this article was, however,
the empirical determination of the characteristic magnitudes of the
in-medium pion and rho meson parameters
that should result from a theoretical model
in order to explain the present experimental data.

The findings of our analysis may be carried over to a prediction of dilepton
spectra for the conditions at RHIC, where our pion annihilation model
together with the QCD phase transition and the suggested in-medium
modifications of pion and rho meson properties will be appropriate.



\end{document}